
 \input harvmac
\input epsf.tex
\def\caption#1{{\it
	\centerline{\vbox{\baselineskip=12pt
	\vskip.15in\hsize=4.2in\noindent{#1}\vskip.1in }}}}
\def\du{{{\rm d} \over {\rm du}}}
\def\uint{\int_0^1 {\rm du\ }}
\def\zint{\int_0^{M} {\rm d}z}

\def\pintv{\int[{\rm d}\Psi][{\rm d}\bar\Psi]\,}

\def\pyidk{PHY-9057135}
\def\CA{{\cal A}}   
   \def\CH{{\cal H}}
   \def\CL{{\cal L}}
  \def\CO{{\cal O}} 
  \def\CS{{\cal S}} 
  \def\CW{{\cal W}} 
 \def\CZ{{\cal Z}}
\def\hatA{\hat\CA}
%
%

\def\bar#1{\overline{#1}}

\def\bra#1{\left\langle #1\right|}
\def\ket#1{\left| #1\right\rangle}
\def\braket#1#2{\bra{#1}#2\rangle}

\def\half{{\textstyle{1\over2}}} 
\def\frac#1#2{{\textstyle{#1\over #2}}} 
%
%
%

\def\Tr{\mathop{\rm Tr}}
\def\Im{\mathop{\rm Im}}

%
%
\def\ltap{\ \raise.3ex\hbox{$<$\kern-.75em\lower1ex\hbox{$\sim$}}\ }
\def\gtap{\ \raise.3ex\hbox{$>$\kern-.75em\lower1ex\hbox{$\sim$}}\ }
\def\roughly#1{\raise.3ex\hbox{$#1$\kern-.75em\lower1ex\hbox{$\sim$}}}
%
%

\def\etal{\hbox{\it et al.\ }}

\def\np#1#2#3{{Nucl. Phys. } B{#1} (#2) #3}
\def\pl#1#2#3{{Phys. Lett. } {#1}B (#2) #3}
\def\prl#1#2#3{{Phys. Rev. Lett. } {#1} (#2) #3}
\def\physrev#1#2#3{{Phys. Rev. D} {#1} (#2) #3}

\relax
\def\Dsl{\,\raise.15ex \hbox{/}\mkern-13.5mu D}
\def\frac#1#2{{\textstyle{#1 \over #2}}}
\def\[{\left[}
\def\]{\right]}
\def\({\left(}
\def\){\right)}

\noblackbox

\Title{
\vbox{\hfill DOE/ER/40561-232-INT95-00-106\smallskip \hfill
UW/PT-95-15\smallskip \hfill BUHEP-95-29}}
{\vbox{\centerline{Domain Wall Fermions}\vskip.2in\centerline{
and the $\eta$-Invariant}}}
\smallskip
\bigskip
\centerline{David B. Kaplan\footnote{$^a$}{{\tt
dbkaplan@phys.washington.edu}}
}
 \centerline{{\sl
Institute for
Nuclear Theory, Box 351550}}
\centerline{{\sl University of Washington, Seattle WA 98195-1550, USA}}
\medskip
\centerline{{\it and}}
\medskip
\centerline{{ Martin Schmaltz\footnote{$^b$}{{\tt schmaltz@bu.edu}}
}}
 \centerline{{\sl
Department of Physics, Boston University}}
\centerline{{\sl 590 Commonwealth Ave., Boston MA 02215, USA}}

\vfill
{  We extend work by Callan and Harvey and show  how the
phase of the   chiral fermion determinant in four dimensions is
reproduced by zero modes bound to   a domain wall in five
dimensions. The analysis could shed light on the applicability
of zero mode fermions and the vacuum overlap formulation of
Narayanan and Neuberger for  chiral gauge theories on the
lattice.}
\Date{10/95}
\baselineskip 18pt
\newsec{Introduction}

In ref. \ref\charv{C.G. Callan, Jr. and J.A. Harvey, \np{250}{1985}{427}.}
Callan and Harvey analyzed fermion zero modes in background gauge fields
bound
to domain wall and vortex defects in arbitrary numbers of dimensions.  The
case
of the domain wall is particularly interesting, since an effective chiral
theory in even dimensions can be embedded in an odd dimensional Dirac
theory.
The purpose of  ref. \charv\ was to elucidate how anomalies of the effective
theory arose due to Chern-Simons currents in the full theory.  It was
subsequently shown that the same system could be implemented on the
lattice
without encountering doubling  of the chiral mode spectrum \ref\dki{D.B.
Kaplan, \pl{288}{1992}{342}.};    Chern-Simons current flow occurs on the
lattice as well \nref\dkii{M.F.L. Golterman, K. Jansen, D.B. Kaplan,
\pl{301}{1993}{219}.}\nref\jns{K. Jansen, M. Schmaltz,
\pl{296}{1992}{374}.}\refs{\dkii,\jns} giving rise to an anomalous  current
divergence in the effective theory along the defect.  The discussion motivated
the vacuum overlap formulation of chiral gauge theories on the lattice of
Narayanan and Neuberger \ref\nn{R. Narayanan and H. Neuberger,
\pl{302}{1993}{62}; \np{412}{1994}{574}; \prl{71}{1993}{3251};
\np{443}{1995}{305}.} who proposed a specific ansatz for the chiral fermion
determinant both on the lattice and in the continuum as the overlap of two
particular quantum states.  The vacuum overlap formalism was subsequently
used
to compute the nonabelian anomaly in four dimensions, using a lattice
regulator
\ref\rds{S. Randjbar-Daemi and J. Strathdee, \pl{348}{1995}{543};
\np{443}{1995}{386}.}
as well as the Lorentz anomaly in two-dimensional quantum gravity
\ref\rdsii{S. Randjbar-Daemi and J. Strathdee, \physrev{D51}{1995}{6617}.}.

Most discussions of the domain wall system have focussed on the anomaly
as a test of the chiral nature of the effective theory (for a recent
exception, see \ref\rgeq{S. Randjbar-Daemi, J. Strathdee, hep-th/9510067}).
In this Letter we return to
the continuum to extend the Callan Harvey results beyond the anomaly to
include
the complete phase of the chiral fermion determinant in gauge backgrounds of
trivial topology.  In particular we show that in taking care to properly
regulate the theory, one can reproduce the $\eta$-invariant formulation of the
chiral phase as developed by Alvarez-Gaum\'e \etal\ \ref\agd{L.
Alvarez-Gaum\'e, S. Della Pietra and V. Della Pietra, \pl{166}{1986}{177};  S.
Della Pietra, V. Della Pietra and L. Alvarez-Gaum\'e, Commun.Math.Phys.109
(1987) 691; L. Alvarez-Gaum\'e and S. Della Pietra, {\it talk presented at
Niels Bohr Centennial Conf.}, published in Copenhagen Bohr Symp. (1985)
95.}
(see also
  \ref\agg{L. Alvarez-Gaum\'e and P. Ginsparg, \np{243}{1984}{449};
Ann.Phys.161 (1985) 423; erratum, ibid.171 (1986) 233.}).  Although we will
restrict our attention to the continuum, it is our hope that the analysis can
eventually shed light on lattice chiral gauge theory.

\newsec{A continuum path integral for chiral fermions}

In four Euclidian dimensions the determinant of a Dirac fermion coupled to
background gauge fields $\det(i\Dsl)$ can be defined as a positive definite
object (by using Pauli-Villars regulators, for example). The determinant of a
chiral fermion $\det(i\Dsl P_L)$ is formally given by
\eqn\phasedef{\det(i\Dsl P_L)\equiv Z[A] \equiv e^{-W[A]}=
e^{-i\phi[A]}\sqrt{\det(i\Dsl)}}
Thus it is the phase $\Im W[A]$  which
 we wish to understand; it contains all of the anomalies in the
theory as well as other physics.   In this section we will try to establish the
connection between the phase $\Im W[A]$ and the phase of the  determinant
for a
five-dimensional Dirac fermion in the presence of a domain wall.

Consider the  Euclidian path integral $\CZ[\CA]$ corresponding to a fermion in
five dimensions with a space dependent mass in a background
gauge field $\CA$ that has the particular form of a four-dimensional gauge
field embedded in five dimensions:
\eqn\vgauge{\CA_\mu = (A_i(x), 0)\ .}
If the mass $m$ only depends on the fifth coordinate $s$, then
\eqn\zfive{\CZ[\CA] = {\rm e}^{-\CW[\CA]} =  \pintv {\rm exp}\(- \CS[\CA]\)}
where
\eqn\sfive{ \CS[\CA] = \int {\rm d}^5x\, \bar \Psi K[\CA]\Psi\ ,\qquad
K[\CA] = i\gamma_5\partial_s - i m(s) + i\Dsl_4\  .}
 In the above expression  $\Dsl_4$ is the four
dimensional covariant derivative, independent of $s$. Four- and
five-dimensional tensor indices are denoted by  Roman and Greek letters
respectively.  The letters $\CA$, $\CS$, $\CZ$ and $\CW$ refer to five
dimensional gauge field, action, partition function and effective action
respectively, while $A$, $S$, $Z$, $W$ denote their four-dimensional
analogues.
We take $\CA$ to
be Hermitian and $\{\gamma_\mu,\gamma_\nu\}= 2 \delta_{\mu\nu}$. Capital
$\Psi(x,s)$  denotes a four component Dirac spinor in five dimensions, while
lowercase $\psi(x)$ will be a four component, four dimensional Dirac spinor.

Since the operator $K[\CA]$ is separable, it is convenient to  expand the
$\Psi$ fields in a product basis of the form
\eqn\fexp{\eqalign{
\Psi(x,s) &= \sum_{n} P_L \psi_n(x) b_n(s) + \sum_{n} P_R \psi_n(x)f_n(s)\
,\cr
\bar\Psi(x,s) &= \sum_{n} \bar \psi_n(x)P_R   b^{*}_n(s) + \sum_{n}
\bar \psi_n(x)P_L f^{*}_n(s)\ .\cr}}
where the $\psi_n(x)$ are arbitrary four dimensional Dirac spinors, and
$P_{R,L}= (1\pm\gamma_5)/2$.  The functions $b_n(s)$ and $f_n(s)$ are
taken to
satisfy the eigenvalue equations
\eqn\heig{\eqalign{
\[-\partial_s^2 + m(s)^2 + \dot m(s)\] f_n(s) &= \mu_n^2 f_n(s)\ ,\cr
\[-\partial_s^2 + m(s)^2 - \dot m(s)\] b_n(s) &= \mu_n^2 b_n(s)\ .\cr}}
In the above equation, $\dot m \equiv \partial_s m$ and the spectrum
$\mu_n$
is assumed to be discrete.  For nonzero $\mu_n$, the $b_n$
and $f_n$ are paired;  however, there can be an arbitrary number of unrelated
$b$ and $f$ zero modes as well.

The equations \heig\ can be regarded as the Schr\"odinger equation for a
supersymmetric quantum mechanical system, with supersymmetry generator
$Q =
[\partial_s +m(s)] \gamma_0 P_L$. The functions $b_nP_L$ and $f_nP_R$
correspond to the ``boson'' and ``fermion'' eigenstates of $\{Q,Q^{\dagger}\}$
which are necessarily degenerate when they have nonzero ``energy''
$\mu_n^2$.
The bosonic and fermionic ``vacuum states'' (zero modes) need not be related
however \ref\witten{E. Witten, \np{185}{1981}{513}; \np{202}{1982}{253}.}.

The five dimensional action \sfive\ can be recast in this basis as a four
dimensional action involving an infinite number of flavors:
\eqn\zprod{ S[A] = \int{{\rm d}^4x\,\[
\sum_{k=1}^{n_b} \bar\psi_k i\Dsl_4 P_L \psi_k +
\sum_{k=1}^{n_f} \bar\psi_k i\Dsl_4 P_R \psi_k +
\sum_{n} \bar\psi_n (i\Dsl_4 -i\mu_n) \psi_n \]\ .   }}
where $n_b$ and $n_f$ are the number of bosonic and fermionic zero mode
solutions to eq. \heig\ respectively, and the final sum excludes the zero
modes.
Provided this action can be suitably regulated, we see that it corresponds to
an infinite tower of massive Dirac fermions with mass $\mu_n$, as well as
$n_b$  left handed and $n_f$ right handed chiral fermions.  If we chose
$m(s)$ and the boundary conditions to eq. \heig\ such that the supersymmetry
is
unbroken, we are guaranteed that there will be at least one chiral fermion,
corresponding to the groundstate of the supersymmetric Hamiltonian.

The system Callan and Harvey considered consisted of a step function for the
mass $m(s)$ in infinite volume.  There was a single zero mode in the action
\zprod, and they showed that the anomaly in the zero mode current was
compensated by the Chern-Simons current induced by integrating out the
heavy
fermions in \zprod.  Instead, we wish to use the five dimensional system to
define the full chiral phase $\Im W[A]$. It is impossible to do so with the
system considered in \charv\ which lacks both regulators and  boundary
conditions in the fifth dimension; we are forced to introduce both \foot{Ref.
\charv\ ignored the regulators  since the induced Chern-Simons current is
finite;  however one finds that the regulators needed to make sense out of the
rest of the theory make finite contributions to the current.  See \dkii\ for
example.}. We will  work in finite volume and use conventional Pauli-Villars
regulators, which requires an action that has an equal number of right and left
handed fields.      In particular, we choose the mass function $m(s)$ to
represent a domain wall -- anti-wall pair with periodic boundary conditions on
the $b$ and $f$ eigenfunctions; the result is a theory with a discrete spectrum
and  a pair of exact zero mode solutions to eq. \heig, $f_0$ and $b_0$ pictured
in fig. 1.
\topinsert{\centerline{\epsfxsize=2.0in\epsfbox{
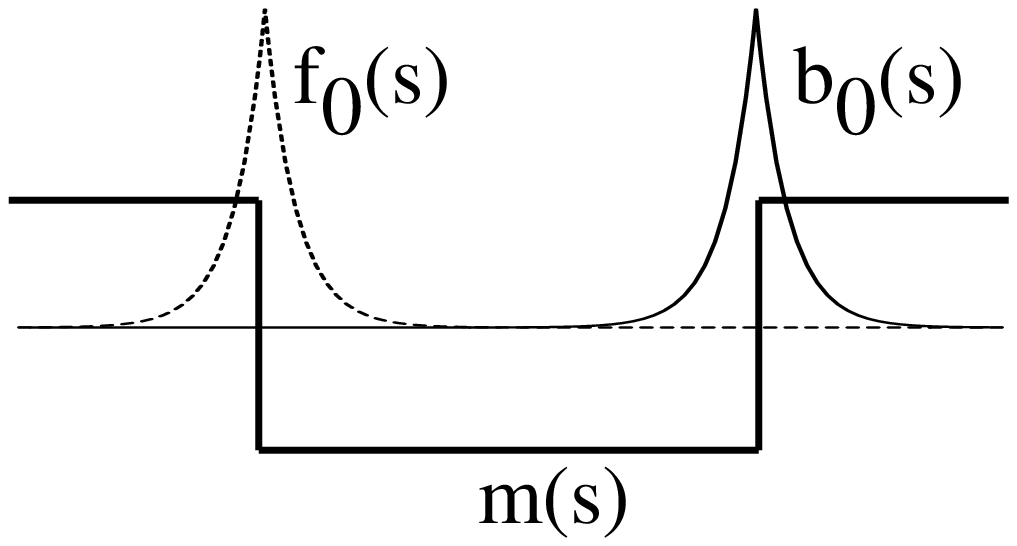}}
\smallskip
\caption{
Fig. 1.  A domain
wall/anti-wall pair at $s=\pm L_s/2$ respectively with periodic boundary
condtions at $s=\pm L_s$ for the functions $b(s)$ and $f(s)$ in eq. \heig.  The
lowest eigenstates are
a pair   of exact zero modes, with the boson localized at $s=L_s/2$
(solid
line) and the fermion localized at $s=-L_s/2$ (dashed line).
}}\endinsert

Wave  function renormalization diagrams for gauge bosons in five dimensions
are
linearly divergent, but since there is no subleading logarithmic divergence
only one Pauli-Villars field is required to regulate the effective action
$\CW[\CA]-\CW[0]$.  We take the regulator to have mass $(m(s)+M)$ and
loop
factor $+1$, where a normal fermion has loop factor $-1$.  The regulated
fermion determinant takes the form
\eqn\regedet{(\det K)_{reg.} = {\det K \over \det(K-iM) }\ .}
We will be examining the phase of this object in the limit that the regulator
mass  $M$ is taken to be large.

Until now we have only considered gauge fields $\CA$ independent of the
coordinate $s$.  However, with $s$-independent gauge fields, both
zero modes
$f_0$ and $b_0$ couple equally, and the theory we are describing is
vector-like
and incapable of reproducing the four-dimensional chiral phase.  This is the
price
of introducing conventional Pauli-Villars regulators.
However one can take advantage of the fact that the two zero modes $f_0$
and
$b_0$  are spatially separated in the fifth dimension, and only couple one of
them to the gauge field.  This is  accomplished by modifying eq. \vgauge,
coupling the fermions to a particular five-dimensional gauge field
$\hat\CA_\mu$ of the form
\eqn\grho{\hat\CA_i = A_i(x)\rho(s)\ ,\qquad \hat\CA_5=0\ ,}
where $\rho(s)$ is a real function of $s$ with $\rho(s)=1$ in the region
$|s-L_s/2|<L_s/2$, smoothly going to zero outside that region.  The region
where
$\rho=1$ is chosen to include the massless ``boson'' state at $s=L_s/2$
 as pictured in fig. 2, but not to overlap with the massless ``fermion'' state
at $s=-L_s/2$.

\topinsert{\centerline{\epsfxsize=2.0in\epsfbox{
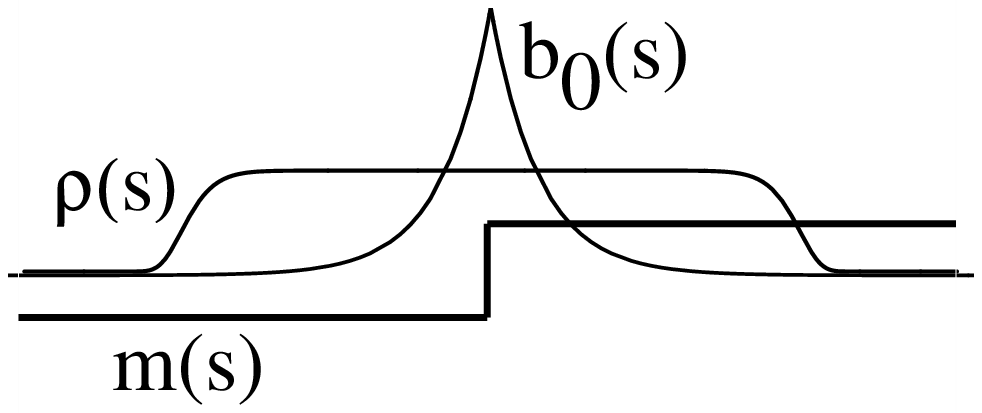}}
\smallskip
\caption{
Fig. 2. Profile in the fifth dimension of the mass function $m(s)$ and the
gauge field weighting function $\rho(s)$, which equals one for
\hbox{$|s-L_s/2|<L_s/2$} and
goes to zero outside that region.  $\rho(s)$  is chosen so that
it is nonzero  in the support of the chiral mode $b_0(s)$, but vanishes in the
vicinity of the mode $f_0(s)$.}}\endinsert

Since $\hatA$ is $s$-dependent,  the fermion operator $K[\hatA]$ is no
longer
separable, and we must deal directly with the regulated five dimensional
effective action $\CW[\hatA]$  rather than treating the theory as  four
dimensional
theory with an infinite tower of flavors.  The central assertion of this paper
is that the phase $\Im W[A]$ of the chiral determinant in four dimensions with
gauge field $A_i(x)$ is given by
\eqn\assert{\Im W[A] -\Im W[0] = \Im \CW[\hatA]-\Im \CW[0]}
where $\hatA$ is the particular five dimensional gauge field given in eq.
\grho.  We first show that eq. \assert\ correctly reproduces the anomaly {\it
\`a la} Callan-Harvey \charv, and then we prove that in fact it reproduces the
complete phase as expressed in papers by Alvarez-Gaum\'e \etal

\newsec{Gauge invariance and the anomaly}

In this section we give a heuristic argument along the lines of ref. \charv\
for why  our expression for the four dimensional chiral
determinant \assert\  correctly reproduces the anomalous phase under four
dimensional gauge transformations.  A more rigorous argument follows in the
next section.

Define the infinitesimal four- and five-dimensional gauge transformations
$\delta_4$ and $\delta_5$:
\eqn\gtiv{\delta_4 A_i(x) = -D_i\theta(x)=-(\partial_i\theta +
\[A_i,\theta\])\
,}
\eqn\gtv{\delta_5 \CA_\mu(x,s) = -D_\mu v(x,s)=-(\partial_\mu v+
\[\CA_\mu,v\])\ .}
Gauge invariance of the five-dimensional theory implies
\eqn\ginv{\delta_5 \Im \CW[\CA] =0\ .}
We wish to prove  that when we vary our ansatz $W[A]$ under a
four-dimensional gauge transformation, we get the correct consistent anomaly
for a four-dimensional chiral fermion \agg:
\eqn\anomi{\delta_4 \Im W[A]\equiv\delta_4 \Im\CW[\hatA] = {1\over
24\pi^2}\int{\rm d}^4x {\rm Tr}
\[\theta {\rm d}(A {\rm d} A +\half A^3)\]\ .}
The reason why five-dimensional gauge invariance in eq. \ginv\ does not imply
four-dimensional gauge invariance of $W[A]$ is that
with $\hat \CA = (A_i(x)\rho(s),0)$, $\delta_4 \hat
\CA\equiv (\delta_4 A_i \rho,0)$ cannot be written as $\delta_5 \hat\CA$.

Inspired by ref. \charv\ we write  the effective action for an arbitrary
five-dimensional gauge field $\CA$  as the sum of  contributions from the
chiral mode plus the contributions from everything else --- the heavy modes:
$$ \CW[\CA] =  \CW_\chi[\CA] + \CW_h[\CA]\ .$$

{}From fig. 2 we see that the chiral mode is localized near the domain wall
at $s=L/2$\ where $\rho=1$ and $\dot\rho=0$. Over that region the four
dimensional gauge transformation $\delta_4$ is identical to a five dimensional
gauge transformation $\delta_5$ with $v(x,s)=\theta(x)\rho(s)$,
therefore
\eqn\delchi{\delta_4 \Im\CW_\chi[\hatA] = \delta_5 \Im\CW_\chi[\hatA]
\Bigl.\Bigr\vert_{v(x,s)=\theta(x)\rho(s)} =
-\delta_5 \Im\CW_h[\hatA]\Bigl.\Bigr\vert_{v(x,s)=\theta(x)\rho(s)}}
where the last equality follows from five dimensional gauge invariance \ginv.
This allows us to express $\delta_4 \Im\CW[\hatA]$ in terms of
$\Im\CW_h[\hatA]$ alone
\eqn\delw{\eqalign{
\delta_4 \Im W[A] =\delta_4 \Im\CW[\hatA] &= \delta_4 \Im\CW_\chi[\hatA]
+ \delta_4 \Im\CW_h[\hatA] \cr
&=  - \delta_5 \Im\CW_h[\hatA]\Bigl.\Bigr\vert_{v=\theta\rho} +
\delta_4 \Im\CW_h[\hatA]
\ .}}
We now determine the two terms separately, using an explicit expression for
$\Im\CW_h[\hatA]$ which we calculate perturbatively in the adiabatic limit,
following \charv.
In the limit of large domain wall mass and smooth gauge fields the lowest
dimensional operator in the adiabatic expansion is
\foot{See also
refs.\nref\naculich{S.G. Naculich, \np{296}{1988}{837}.}\nref\schand{S.
Chandrasekharan, \physrev{49}{1994}{1980}.} \refs{\naculich-\schand}\ for
more
detailed computations than found in ref. \charv.}
\eqn\wheavy{\Im\CW_h[\CA]=-\pi \int {\rm d}^5x \( {m(s)\over |m(s)|} - {M\over
|M|}\) Q^0_5[\CA]}
where $Q^0_5$ is the Cern-Simons form
\eqn\CSform{Q_5^0[\CA] = {1\over (2\pi)^3 2!}\int_0^1 {\rm d}\sigma\,
\Tr\[\CA(\sigma {\rm
d}\CA +  \sigma^2 \CA^2)^2\]\ .}
It is easily verified by substitution that
$$\Im\CW_h[\hatA]=0  \qquad {\rm and} \qquad
\delta_4 \Im\CW_h[\hatA]=0$$
for the gauge field configuration $\hatA$ defined in \grho.
Calculating the $\delta_5$ term in \delw\ yields
\eqn\dwh{\eqalign{-\delta_5 &\Im\CW_h[\hatA]
\Bigl.\Bigr\vert_{v=\theta\rho}\cr &={1\over 24\pi^2}
\int{\rm d}^4x\,{\rm d}s\,\(\delta(s-{L_s\over2})-\delta(s+{L_s\over2})\) {\rm
Tr} \[v
{\rm d}(\hatA {\rm d} \hatA +\half \hatA^3)\]
\Bigl.\Bigr\vert_{v=\theta\rho} \cr
&={1\over 24\pi^2}\int{\rm d}^4x {\rm Tr} \[\theta {\rm d}
(A {\rm d} A +\half A^3)\]\ .}}
In the final step above we used the fact that $\rho(L_s/2)=1$,
$\rho(-L_s/2)=0$.
Substituting this result and $\delta_4 \Im\CW_h[\hatA]=0$ into \delw\  we
finally obtain  the result we set out to prove in eq. \anomi\ --- namely, that
the variation of our ansatz \assert\ for $W[A]$ under four-dimensional gauge
transformations correctly reproduces the consistent non-Abelian anomaly for
a
chiral fermion.

\newsec{The complete chiral phase and the $\eta$ invariant}

We now arrive at the central point of this paper, which is to show that $\Im
W[A]$ as defined in eq. \assert\ reproduces not just the anomaly, but the
entire phase of the chiral determinant (in the zero instanton sector).  In
refs. \refs{\agd,\agg}\ it was shown that the  chiral phase  can be related to
certain properties of the
five-dimensional operator
\eqn\hag{\CH= i\partial_s \gamma_5+ i\Dsl[\CA_i]}
where $\CA_5=0$ and $\CA_i(x,s)$ describes a path in the space of
four-dimensional gauge fields
from  $\CA_i(x,-\infty)=\bar A_i(x)$  to $\CA_i(x,+\infty)=A_i(x)$. The
four-dimensional gauge field $\bar A(x)$ plays the role of some  fiducial gauge
field which is left fixed, while $A$ can be varied.  In particular, ref. \agd\
showed that in the zero instanton sector, the effective action for a chiral
fermion in four dimensions may be expressed as
\eqn\agresult{\Im(W[A] - W[\bar A]) = \pi (\eta[\CH]+{\rm dim\  Ker}\ \CH) -
2\pi Q_5^0[\CA_i]\ .}
$Q_5^0$ is the Chern-Simons form  given in eq. \CSform;
 $\eta[\CH]$ is the so-called $\eta$-invariant of  $\CH$,  defined to be the
regulated sum of the signs of its eigenvalues,  $\sum \lambda/|\lambda|$,
where possible zero modes are omitted in this definition.  The contribution of
 $\CH$ zero modes  are accounted for by the
 $({\rm dim\ Ker}\ \CH)$ term. The
combination of $\eta$ and $Q_5^0$ is independent of the path $\CA$, and
only depends on the endpoints $\bar A$ and $A$ \ref\aps{M.F. Atiyah, V.K.
Patodi and I.M. Singer, Math. Proc. Camb. Phil. Soc. 77 (1975)
 43; 78 (1975) 405; 79 (1976) 71.}.
The authors of ref. \agd\ further showed that the $\eta$-invariant  can be
expressed as
\eqn\reget{\eqalign{
\eta[\CH] &= -{1\over\pi}\lim_{M\to\infty} \Im\Tr\ln\[{\CH-i
M\over \CH+iM}\] \cr &=  -{1\over\pi}\lim_{M\to\infty} \lim_{L\to\infty}
 \uint \du \Im\Tr P_L \ln\[{\CH^u-iM\over \CH^u+iM}\]\ ,}}
where $P_L(s)=\theta(s+L)\theta(L-s)$, and  zero
modes of $\CH$ are omitted.
$P_L$ is inserted to keep track of the
noncompactness of the manifold by providing explicit boundaries which are
removed to infinity.
The object $\CH^u$ has been introduced so that the
endpoint of the path in gauge field space can be smoothly changed as a
function
of the parameter $u$: $\CH^u = \CH[ \CA_i^u]$ with $\CA_i^u(x,-\infty)=\bar
A_i(x)$ and $\CA_i^u(x,+\infty)=A_i^u(x)$ with $A_i^{u=0}(x)=\bar A_i(x)$
and
$A_i^{u=1}(x)=A_i(x)$.

We now demonstrate that the path integral for domain wall fermions (eq.
\assert) can be rewritten in the form of the right side of eq. \reget, thereby
reproducing the result \agd\ of Alvarez-Gaum\'e \etal To
achieve this, we use the Pauli-Villars regulated expression  \regedet\ for the
five-dimensional determinant and consider the particular family of
paths in field space of the form
\eqn\ourpaths{\eqalign{\hat\CA_\mu^u(x,s) &= (A^u_i(x)\rho(s)
+\bar A_i(x)(1-\rho(s)),0)\ ,\cr
A_i^{u=0}&=\bar A_i(x)\ ,\qquad A_i^{u=1}(x) = A_i(x) }}
where $\rho(s)$ is the smooth function discussed previously and pictured
in fig. 2.
For the  wall/anti-wall configuration pictured in fig. 1 we can now write $\Im
\CW$ as
\eqn\imw{\eqalign{\Im (\CW[\hatA^{u=1}] - \CW[\hatA^{u=0}])
        &= \Im \uint \du \( \ln {\rm det}
         \[{K[\hat\CA^u] \over K[\hat\CA^u]-iM}\] \) \cr
	&=  \uint \du \Im \Tr \[
         (P_I+P_{II}+P_{III}) \ln {K[\hat\CA^u] \over K[\hat\CA^u]-iM
}\]}}
in the appropriate limits that the fermion and Pauli-Villars masses and the box
size $L_s$ are taken to infinity.
To compute the integral in $\imw$  we
have divided the fifth dimension into four regions, as shown in fig. 3.
This is done  by inserting $1=P_I+P_{II}+P_{III}+P_{IV}$ into the trace
defining the effective action $W$, where   $P_R=1$  in region $R$ and
$P_R=0$ elsewhere.    Region IV includes the ``fermionic'' mode bound to the
anti-domain wall (not pictured), corresponding
to a right-handed chiral fermion, and $\rho(s)=0$
in this region.  Thus $\frac{\rm d}{\rm du}\hat\CA_\mu^u=0$  and so region IV
does not contribute to
the expression $\imw$.
\topinsert{\centerline{\epsfxsize=2.0in\epsfbox{
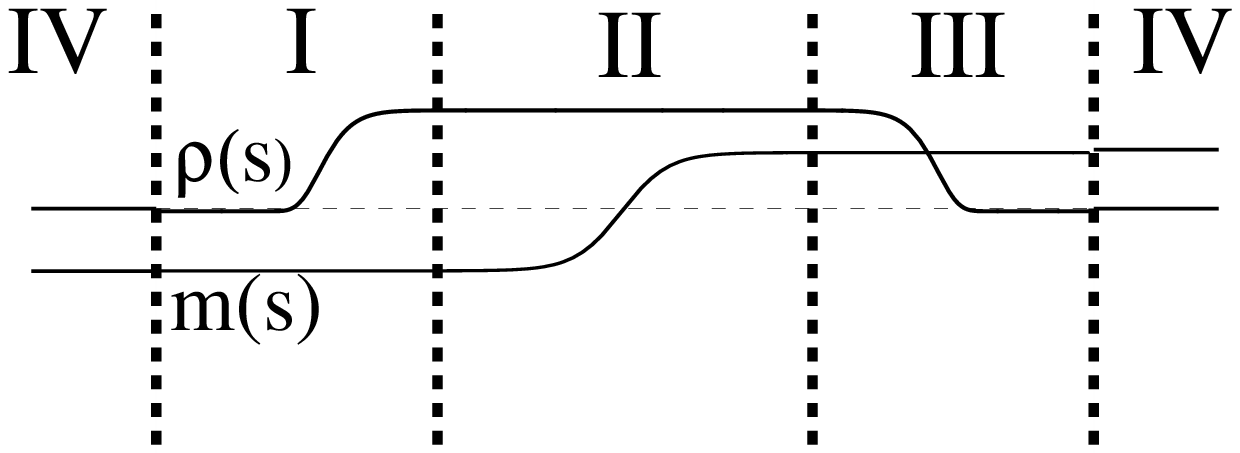}}
\smallskip
\caption{
Fig. 3. The fifth dimension is divided into four regions for the purpose of
computing $\Im W[A]$ in eq.\imw.  The divisions occur far from where either
$\rho(s)$ or $m(s)$ are varying. The gauge field \ourpaths\ is independent of
$u$ in region IV and therefore that region does not contribute. }}\endinsert

We now compute the contributions to the trace from each of the  regions
I--III.
Note that our operator $K[\CA^u]$ may be written in terms of the
Alvarez--Gaum\'e \etal\ Hamiltonian \hag\ as
$$K[\CA^u] = \CH^u - im(s)\ .$$
In regions I and III the step function mass $m(s)$ equals $\mp m_0$ so
that $K[\CA^u_i]= \CH^u\pm i m_0$ respectively, and
note that the paths \ourpaths\ -- restricted to regions I and III -- are
examples of the paths used in the expression for the $\eta$ invariant \reget.
In region I we have
\eqn\regI{\eqalign{
	\Im &\Tr \(P_I \ln  {K \over K-iM}\)\cr
	&=-{i \over 2} \Tr P_I
	\( \ln  {\hat \CH^u+i m_0 \over \hat \CH^u-i m_0}
	-  \ln  {\hat \CH^u+i (m_0-M) \over\hat \CH^u-i (m_0-M)}\)\ .}}
Comparing the above expression with the expression \reget\
for the $\eta$-invariant, we find
\eqn\regIa{\eqalign{\lim_{m_0\to\infty} \lim_{M\to \infty}\lim_{L_s\to\infty}
&\uint \du
	\Im \Tr \(P_I  \ln  {K \over K-iM }\) \cr =&\
      (1+1){\pi \over 2} (\eta[\CH]+{\rm dim\ Ker}\ \CH)  =
           \pi(\eta[\CH]+{\rm dim\ Ker}\ \CH)\ , }}
where we have made explicit the equal contributions from the fermion  and
Pauli-Villars fields.
Note the ordering of limits:  first we must send the
wall/antiwall separation to infinity;  then we take the regulator mass to
infinity; and finally we take the domain wall height $m_0$ to infinity. This
ensures that first the interaction between the gauge field and the unwanted
zero mode at $s=-L_s/2$ is sent to zero; then the Pauli-Villars fields are
decoupled from the five
dimensional theory; and finally the heavy five-dimensional fermion modes are
decoupled from the effective four dimensional theory on the domain wall.

The signs of the fermion and Pauli-Villars contributions to the above
expression are determined
from \regI\ by the signs of the masses; thus in region III where the step
function is positive, the fermion contribution changes sign and the two fields
give a combined contribution
$\propto (-1+1)=0$  --- there is no contribution from this region to $\Im W[A]$
\foot{This behaviour is directly related to the fact that when the theory is
regulated, the Chern-Simons current flow only comes from region I (with twice
the value found in ref. \charv) --- see eq. \wheavy.}.

The contribution from region II --- where the fermion mass is changing, but the
gauge field is constant --- vanishes as well. To see this we write the
contribution as
\eqn\regII{\Im\Tr \( P_{II} \ln  {K\over K-iM }\)
      = \Im \zint \Tr \( P_{II} {i\over K-iz} \)\ .}
But the operator $\CO\equiv  i\Dsl\gamma_5$ satisfies $\CO K \CO^{-1} = -
K^{\dagger}$
and $\[\CO,P_{II}\]=0$; thus
\eqn\regIIreal{\eqalign{\Im \Tr \( P_{II} {i\over K-iz}\) &=
\Im \Tr \(\CO\CO^{-1} P_{II} {i\over K-iz}\) \cr &=
\Im \Tr  P_{II} \({i\over K-iz}\)^{\dagger} \cr &=0\ .}}

It follows that the five-dimensional path integral for domain wall fermions in
the  sequential limits of  eq. \regIa\ has the phase
\eqn\ourphase{\Im(\CW[\hatA^{u=1}] - \CW[\hatA^{u=0}])=   \pi(\eta[\CH]
+{\rm
dim\ Ker}\ \CH)}
where $\CH$ is the operator \hag\ as a functional of the gauge path
$\hat\CA_\mu=(A_i\rho+\bar A_i(1-\rho),0)$ acting in region I in fig. 3,
in the limit that the region  is infinite in extent.  The right hand
side of \ourphase\ agrees with the general result of
Alvarez-Gaum\'e \etal\ \agresult, provided that the
Cern-Simon form $Q_5^0[\hatA]$  vanishes.  If we
restrict ourselves to paths $\hat A$ with
 $\bar A=0$, then $Q_5^0[\hatA]=0$ and we have
proven our central assertion \assert\ that five dimensional domain
wall fermions with $\bar A=0$ correctly reproduce the complete chiral
phase in the zero instanton sector. (Note that $\bar A=0$ implies
$\hatA^{u=0}=0$).

It would be very interesting to pursue this analysis beyond the
zero instanton sector, but we do not do so here.

Finally, we mention that the domain wall formula correctly reproduces the
SU(2) Witten anomaly \witten.   Consider an $SU(2)$ gauge theory with an odd
number of $SU(2)$-doublet Weyl fermions. For real and pseudo-real
representations $Q_5^0[\hat A]=0$, and so the domain wall result \ourphase\
reproduces the exact result \agresult\ for any gauge field $\bar A$. If the
gauge field $\bar A = A^g$ is related to $ A$ by a large gauge transformation
$g$  which is
a nontrivial element of $\pi_4(SU(2))$, then by the Atiyah-Singer theorem,
(${\rm dim Ker}\CH$) is an odd integer. Furthermore, $\eta(\CH)$  vanishes
because for real (and for pseudo-real) representations the nonzero eigenvalues
of $\CH$ come in opposite sign pairs \agd. Thus our formula \ourphase\
correctly reproduces Witten's anomaly \witten $$ \Im W[A]=\Im W[A^g]+ \pi\
,$$ and the fermion determinant picks up a minus sign under the large
nontrivial gauge transformation.

\newsec{Relation to the vacuum overlap formulation}

 Motivated by domain wall fermions, Narayanan and Neuberger have
suggested that
one can write
 $\det(i\Dsl P_L)$ as  the overlap between ground
states of different Hamiltonians:
\eqn\nndet{{\det(i\Dsl P_L)\over \det(i\dsl P_L)}= {\braket{A -}{A +}\over
\braket{0 -}{0 +}}}
where the states $\ket{A \pm}$ refer to the ground states of the two different
four-dimensional Hamiltonians $H^{\pm}$  in the same background gauge field
$A_i(x)$:
 $$H^{\pm}[A] = i\Dsl_4[A] \mp im_0$$ in the limit $m_0\to\infty$, with the
phase convention that $\braket{0+}{A+}$ and $\braket{0-}{A-}$ are real \nn.
The correct
anomalous  transformation  of this phase has been computed in the
continuum
(without regulators) in $1+1$ dimensions for an Abelian theory  by Narayanan
and Neuberger \ref\nnii{R. Narayanan and H. Neuberger,
\pl{348}{1995}{549}.},
and for a non-Abelian theory in $3+1$ dimensions by Randjbar-Daemi
and Strathdee \rds.

What we will now show is that for the zero instanton gauge backgrounds that
we
are considering in this Letter, the vacuum overlap expression \nndet\
reproduces not just the anomaly, but the complete phase of the chiral
determinant.  For such gauge fields, $\ket{A\pm}$ can be written as \nn
\eqn\nnlim{\lim_{S\to \infty } N e^{-S H^{\pm}}\ket{0\pm}}
where the normalization constant $N$ depends on both $S$ and $A$, but is
real.
Thus the chiral phase predicted by the overlap formulation can be written in
path integral form as
\eqn\nnphase{\Im W[A]-\Im W[0] = \pintv
e^{-\int_{-\infty}^{\infty} {\rm d}s\,\int {\rm d}^4x\, \CL[A]}}
where $\CL[A]$ is the Lagrange density
\eqn\nnlag{\CL[A] = \bar\Psi(x,s) \[ i\gamma_5\partial_s +i\gamma_j(\partial_j
+ iA_j(x)\bar\rho(s)) -i\bar m(s)\]\Psi(x,s)}
with a step function mass $\bar m(s)=m_0\epsilon(s)$, and a gauge function
$\bar\rho(s)$ which is of the form
\eqn\nnrho{\bar\rho(s) = \lim_{s_0\to\infty} \theta(s_0+s)\theta(s_0-s)\ .}

Evidently the phase of the vacuum overlap formula in the zero instanton
sector
is a special case of our formula for domain wall fermions derived in the
previous section, involving a particular choice for the functions $m(s)$ and
$\rho(s)$.  It follows that the ansatz \nndet\ also correctly reproduces the
full chiral phase. The appeal of the overlap formulation is that it can be
implemented on the lattice in a straightforward fashion, and that it can be
used to calculate Green functions in gauge backgrounds of nontrivial
topology.

\newsec{Conclusions}

We have shown how to relate the phase of the chiral fermion determinant in a
background non-Abelian gauge field to the phase of a five-dimension path
integral over Dirac fermions interacting with a domain wall.  In particular, we
have demonstrated how the $\eta$-invariant description of the chiral phase
due
to
Alvarez-Gaum\'e \etal\ arises when the theory is properly regulated with well
defined boundary conditions on the fields.  The phase we have calculated
includes not only the perturbative anomaly, but also possible Witten
anomalies
\ref\wittenii{E. Witten, \pl{117}{1982}{324}.}, as well as non-anomalous
contributions.  The result was seen to extend to the vacuum
overlap formulation of chiral fermions.

A serious deficiency in our arguments is that they only apply to the
topologically trivial gauge sector; it would be interesting if the arguments
could be extended to discuss the 't Hooft vertex in the presence of instantons.
 While the
domain wall model allows one to compute the chiral phase, it is less evident
how to use it to define fermion Green functions for a chiral theory.

Application of these ideas to the calculation of the chiral phase in a lattice
regulated theory is promising since
the periodic boundary conditions we employed are simple to implement  and
fermion doublers can be eliminated by means of a
Wilson term \dki.  The tricky part of a lattice realization of domain wall
fermions lies in the limiting process in eq. \regIa.
A crucial step in the analysis was that the regulator mass $M$ was taken to
infinity with fixed background gauge field, so that  gauge fields and their
derivatives were always small compared to $M$.  On a lattice, the role of $M$
is played  by
the inverse lattice spacing for the fermion fields.    However, in the usual
formulation of lattice gauge theories, the spatial variation of the gauge field
is set by the same lattice spacing, and so the limiting process needed for
reproducing the correct phase is not obtainable \foot{One might think that in
the weak coupling limit, only smooth gauge fields would be present;  however,
gauge symmetry is explicitly broken in the domain wall formulation, as
discussed in \S3, and the fermions are sensitive to the wildly
fluctuating gauge fields  found on every gauge orbit.}.    However, the
problem would seem to be resolved if the gauge fields were integrated over a
coarser lattice than the fermions. The limit $a_\psi/a_U \to 0$ for the ratio
of the lattice spacings is analogous to the $M\to \infty$ limit in the
Pauli-Villars regulated continuum theory.
This suggestion is in line with a  recently proposed solution to the chiral
fermion problem by Hernandez and Sundrum \ref\hersun{P. Hernandez, R.
Sundrum,
preprint hep-ph/9506331; hep-ph/9510328.}.  We
believe that this would also alleviate  problems with the domain wall
formulation
discussed recently in ref. \ref\gosham{M. Golterman, Y. Shamir,
\physrev{D51}{1995}{3026}.}.

Some of the delicate limiting procedures necessary for using domain wall
fermions on the lattice  might be evaded by pursuing the vacuum overlap
formulation of chiral gauge theories, rather than its domain wall progenitor.
It would be useful to better understand the connection between the two
approaches, particularly for topologically nontrivial gauge fields.

\vskip1in
\centerline{Acknowledgements}

We  wish to
thank P. Arnold, L. Brown, R. Narayanan, H. Neuberger, S. Randjbar-Daemi,
and L. Yaffe for useful
conversations.
DK was supported in part by DOE grant DOE-ER-40561, NSF
Presidential Young Investigator award \pyidk, and by a grant from the
Sloan Foundation. During this work MS was supported in part  under
DOE grants DOE-ER-40561 and DE-FG02-91ER40676.

\listrefs
\bye